\documentclass{ws-procs9x6}

\begin{document}

\title{RADIATIVE ELECTROWEAK SYMMETRY BREAKING IN\\
 TeV-SCALE STRING MODELS}

\author{N. KITAZAWA}

\address{Department of Physics, Tokyo Metropolitan University,\\
Hachioji, Tokyo 192-0397, Japan\\
E-mail: kitazawa@phys.metro-u.ac.jp}

\begin{abstract}
We examine the possibility of necessary and inevitable
 radiative electroweak symmetry breaking by
 one-loop radiative corrections in a class of string models
 which are the string realization of ``brane world'' scenario.
As an example,
 we consider a simple quasi-realistic model based on a D3-brane at
 non-supersymmetric ${\bf C}^3/{\bf Z}_6$ orbifold singularity,
 in which the electroweak Higgs doublet fields are identified
 with the massless bosonic modes of the open string on that D3-brane.
We calculate the one-loop correction to the Higgs potential,
 and find that its vacuum expectation value can be realized
 in this specific model.
\end{abstract}


\bodymatter

\section{Introduction and Motivations}

The electroweak symmetry breaking in the standard model
 is not necessary and inevitable.
Though technicolor theories realize
 necessary and inevitable electroweak symmetry breaking,
 like the chiral symmetry breaking in QCD,
 it is not easy to construct simple viable models
 which is consistent with the experimental data
 \cite{PT-letter,PT-full}.
The radiative electroweak symmetry breaking scenario
 in viable models with TeV-scale supersymmetry is also a candidate
 of the necessary and inevitable electroweak symmetry breaking,
 but it requires a certain level of fine tuning
 among parameters in models\cite{KLNW},
 which reduces the necessity and inevitability. 
There are many scenarios utilizing extra space dimensions,
 but it seems very difficult to construct
 theoretically well-defined concrete models
 with necessary and inevitable electroweak symmetry breaking.

Although much efforts have been made
 towards clarify the mechanism of the electroweak symmetry breaking
 in the framework of the quantum field theory,
 there are few efforts in the framework of the string theory,
 in spite of its potential as the unified theory including gravity.
The radiative electroweak symmetry breaking
 in string models using D-branes is first examined
 by Antoniadis, Benakli and Quir\'os\cite{ABQ}.
They have analyzed a concrete string model with no supersymmetry,
 and have shown that
 some tree-level massless scalar fields (open string states)
 can follow non-trivial potential at one-loop level,
 and obtain non-zero vacuum expectation values.
The scale of the vacuum expectation value
 is the radius of some compactified space,
 which is difficult to determine in the framework of
 perturbative (even non-perturbative) string theory.

The author has pointed out the possibility of
 the radiative electroweak symmetry breaking in which
 the scale of the vacuum expectation value is determined
 by the string scale (string tension)\cite{kitazawa}.
This necessary and inevitable electroweak symmetry breaking
 can happen in a class of non-supersymmetric TeV-scale string models
 which is constructed by ``bottom-up approach''\cite{AIQU}.
 
In the following,
 the model of Ref.5 is very briefly introduced,
 and discuss possibility and problems towards constructing
 realistic models.

\section{A Quasi-Realistic Model}

The model is based on ten-dimensional type IIB superstring theory.
The six space dimensions
 are considered to be compactified to some orbifold space
 which is the toroidal compactification with
 some identifications of points by some discrete transformations.
Consider D3-branes
 whose all the three space directions are not compactified.
In six-dimensional compactified space,
 the places of such D3-branes are specified by a point.
In case that D3-branes are at a orbifold singular point
 (fixed point by some discrete transformation),
 it is shown that gauge-interacting chiral fermions
 can be realized in the four-dimensional world-volume
 of the D3-branes,
 and it is possible to construct models of particle physics
 (see Ref.6 and references therein).
Depending on the nature of the singularity,
 the world on D3-branes becomes supersymmetric or non-supersymmetric.

Since many properties of the world on D3-brane
 are determined locally by the nature of the singularity,
 we may only specify the nature of the singularity
 without concretely specify the six-dimensional compactified space
 as the first step.
Consider ${\bf C}^3/{\bf Z}_6$ orbifold singularity,
 where ${\bf C}$ denotes two dimensional space parameterized
 by one complex coordinate and
 ${\bf Z}_6$ denotes a special discrete rotational transformation
 which does not preserve the supersymmetry on the D3-branes
 \cite{kitazawa}.
Only ${\bf Z}_6$ invariant states are survived on the D3-branes.
The original massless open string states on a stack of $N$ D3-branes
 are described by the four-dimensional ${\cal N}=4$
 supersymmetric U$(N)$ Yang-Mills theory.
The ${\bf Z}_6$ projection can break original U$(N)$ gauge symmetry
 by setting non-trivial ${\bf Z}_6$ transformation property
 on the Chan-Paton factor of the open string on the D3-branes.

By considering a stack of $N=6$ D3-brane
 at a ${\bf C}^3/{\bf Z}_6$ singularity
 with some specific ${\bf Z}_6$ transformation properties
 of Chan-Paton indices,
 we have U$(3)\times$U$(2)\times$U$(1)$ gauge symmetry
 with the following massless particle contents on the D3-branes
 (see Ref.5 for detail).
\begin{center}
\begin{tabular}{cccc}
field                 & U$(3)$ & U$(2)$ & U$(1)$ \\
\hline
$q_L \times 3$        & $3$    & $2^*$  & $0$ \\
${\bar u}_L \times 3$ & $3^*$  & $1$    & $1$ \\
\hline
$H \times 3$          & $1$    & $2$    & $-1$  
\end{tabular}
\end{center}
Namely, we have three generations of left-handed quark doublets
and right-handed up-type quarks and three Higgs doublet fields
with the definition of the hypercharge
\begin{equation}
 Q_Y \equiv
  - \left( {{Q_3} \over 3} + {{Q_2} \over 2} + Q_1 \right),
\end{equation}
 where $Q_n$ ($n=1,2,3$) is the U$(1)$ charges of U$(n)$ on D3-brane.
It is clear that this system has chiral anomalies
 and more chiral fermions should be introduced
 to become a consistent system.
In the language of the string theory,
 we have to introduce some D7-branes to cancel
 Ramond-Ramond tadpoles.
Introduction of D7-branes gives further massless particles
 and the chiral anomaly is cancelled out.

Three massless Higgs fields follow the potential
\begin{eqnarray}
 V
   = {{g^2} \over 4}
     \sum_{r,s=1,2,3}
     \left(
      (H^\dag_r H_s)(H^\dag_s H_r)
      +
      (H^\dag_r H_r)(H^\dag_s H_s)
     \right)
\label{potential}
\end{eqnarray}
 which has no flat directions ($g$ denotes a gauge coupling).
Therefore,
 if one-loop radiative correction gives negative mass squared,
 Higgs fields necessary have vacuum expectation values.
These Higgs fields also have the following Yukawa couplings.
\begin{equation}
 {\cal L}_Y =
  - g \sum_{r,s,t=1,2,3} \epsilon_{rst} {\bar u}_R^r q_L^s H^t + {\rm h.c.}
\label{Yukawa}
\end{equation}
Though these Yukawa couplings do not have rich structure
 to generate hierarchical quark mass spectrum,
 the mixing of three Higgs doublet fields
 may realize hierarchy.
Three Higgs field is not necessary equivalent at one-loop level
 depending on the way of introduction of D7-branes.
 
\section{One-loop correction to the Higgs Potential}

The one-loop radiative correction to the Higgs mass
 can be calculated in string world-sheet theory.
The two-point function of a Higgs state is divided by two parts
\begin{equation}
 A_{\rm D3-D3}=A^{\rm NS}_{\rm D3-D3}+A^{\rm R}_{\rm D3-D3}
\end{equation}
 where D3-D3 means the contribution of open string on the D3-branes,
 NS means Neveu-Schwarz sector describing boson loop contributions,
 and R means Ramond sector describing fermion loop contributions.
There is a similar contribution by the open string
 whose one edge is on the D3-branes
 and another edge is on the D7-branes.
\begin{equation}
 A_{\rm D3-D7}=A^{\rm NS}_{\rm D3-D7}+A^{\rm R}_{\rm D3-D7}
\end{equation}
Since boson and fermion loop give positive and negative
 contributions to the Higgs mass squared, respectively,
 if the fermionic contribution is larger than the bosonic one,
 the one-loop Higgs mass squared becomes negative.

The calculation is straightforward,
 but some comments must be given before presenting the result.

The one-loop diagram of the open string
 can be understood as the tree propagation of the closed string.
The one-loop correction to the Higgs mass squared
 can be described as the summation of closed string propagation
 from D3-brane to D3- or D7-branes with different transverse momenta
 (transverse momenta, which are not conserved,
  mean the momenta in the direction perpendicular
  to the D3 or D7-branes).
If there exist massless open string states
 which have tadpole coupling to D3-branes,
 they give infrared divergences in the zero transverse momentum limit.
This is so called NS-NS tadpole problem.
In the present model
 NS-NS tadpole problem exists
 in the contribution of untwisted closed string,
 and it is absent in the contribution of twisted closed string.
Twisted closed staring
 is the string closed up to ${\bf Z}_6$ identifications,
 and untwisted closed string is the usual closed string.
(There are two untwisted closed string sector
 and four twisted closed string sector in the present model.)
The existence of NS-NS tadpole problem
 is understood as the signature that the assumed background geometry
 (or vacuum)
 is not the solution and some background redefinition is required
 \cite{FS1,FS2}. 
It is difficult to solve this problem,
 since the formulation of the strings
 propagating non-trivial background is difficult.
It may be possible that
 we can calculate physical quantity even in wrong ground state
 by the method of ``tadpole resummation''\cite{DPNS}.
At present,
 we simply neglect the contributions of untwisted closed string.

There is a tachyon mode in the untwisted open string sector,
 which also means that the vacuum is unstable.
It is expected that
 the model without tachyon mode can be constructed
 by considering special compactification or projection.
Since the origin of the tachyon mode in the present model
 is related with the tachyon mode in type 0B theory
 \cite{Font-Hernandez},
 and since the tachyon mode in type 0B theory is removed
 by a special projection (Sagnotti's type 0'B model
 \cite{Sagnotti}),
 the model without tachyon mode is expected to be constructed
 by using type 0'B model.

The result of the calculation of one-loop Higgs mass squared is
\begin{equation}
 m^2 \sim
 -
 {{g^2} \over {16 \pi^2}}
 {2 \over {\alpha'}}
 {{36\sqrt{3}} \over \pi}
 \int_0^\infty {{ds} \over s}
 {{e^{- 2 \pi^2 / s}} \over {1-e^{- 2 \pi^2 / s}}}
 e^{-s/3},
\end{equation}
 where the contribution of the untwisted closed string sector
 and the physical divergence due to the Yukawa coupling with
 massless fermions (see Eq.(\ref{Yukawa})) are neglected.
This is an order estimate
 by neglecting the exponentially suppressed higher order
 contribution in the twisted open string sectors.
The one-loop contribution to the mass squared is negative,
 because, technically, bosonic one-loop contributions
 of twisted sectors are cancelled out with each other
 due to the non-trivial ${\bf Z}_6$ transformation rule
 of Chan-Paton indices.
The order of the ``electroweak scale'' is numerically obtained as
\begin{equation}
 v \sim \sqrt{-m^2/g^2} \simeq 10^{-2} \alpha'^{-1/2},
\end{equation}
 where we used that the Higgs quartic coupling
 is given by the gauge coupling (see Eq.(\ref{potential})).
This is consistent with the standard expectation
 that the scale is essentially given by the string scale,
 $1/\sqrt{\alpha'}$, with one-loop suppression factor.
This result suggests that
 necessary and inevitable radiative electroweak symmetry breaking
 is possible in this type of models with D-branes
 at non-supersymmetric orbifold singularities.

It would be very interesting
 to explorer realistic models in this direction.
If this scenario would be true,
 we would observe very rich phenomena
 in future high-energy colliders, LHC for example,
 which are very different from that would be expected
 in field theory models\cite{CPP}.

\section*{Acknowledgments}

The author would like to thank A.Sagnotti
 for fruitful discussions and
 the kind hospitality during the stay in
 The Scuola Normale Superiore in Pisa.
This work has been supported in part by
 the Selective Research Fund of Tokyo Metropolitan University.

\bibliographystyle{ws-procs9x6}
\bibliography{kitazawa}

\end{document}